# LinkML: an open data modeling framework

Sierra A.T. Moxon [0000-0002-8719-7760][1*], Harold Solbrig [0000-0002-5928-3071][2], Nomi L. Harris [0000-0001-6315-3707][1], Patrick Kalita [0000-0002-6150-307X][1], Mark A. Miller [0000-0001-9076-6066][1], Sujay Patil [0000-0001-6142-1106][1], Kevin Schaper [0000-0003-3311-7320][3], Chris Bizon [0000-0002-9491-7674][4], J. Harry Caufield [0000-0001-5705-7831][1], Silvano Cirujano Cuesta[5], Corey Cox [0000-0001-9042-5982][3], Frank Dekervel [0009-0008-4913-7127][6], Damion M. Dooley [0000-0002-8844-9165][7], William D. Duncan [0000-0001-9625-1899][8], Tim Fliss [0000-0003-3303-658X][9], Sarah Gehrke [0000-0003-3245-2880][3], Adam S.L. Graefe [0009-0004-8124-8864][10], Harshad Hegde [0000-0002-2411-565X][11], AJ Ireland [0000-0003-1982-9065][1], Julius O.B. Jacobsen [0000-0002-3265-1591][21], Madan Krishnamurthy [0000-0002-9767-3636][3], Carlo Kroll [0009-0008-4562-7399][3], David Linke [0000-0002-5898-1820][13], Ryan Ly [0000-0001-9238-0642][14], Nicolas Matentzoglu [0000-0002-7356-1779][15], James A. Overton [0000-0001-5139-5557][16], Jonny L. Saunders [0000-0003-0545-5066][17], Deepak R. Unni [0000-0002-3583-7340][18], Gaurav Vaidya [0000-0003-0587-0454][4], Wouter-Michiel A.M. Vierdag [0000-0003-1666-5421][19], LinkML Community Contributors, Oliver Ruebel [0000-0001-9902-1984][14], Christopher G. Chute [0000-0001-5437-2545][20], Matthew H. Brush [0000-0002-1048-5019][3], Melissa A. Haendel [0000-0001-9114-8737][3], Christopher J. Mungall [0000-0002-6601-2165][1]

1 Environmental Genomics and Systems Biology, Lawrence Berkeley National Laboratory, Berkeley, CA 94720, USA
2 School of Medicine, Johns Hopkins University, Baltimore, MD 21287, USA
3 Department of Genetics, University of North Carolina at Chapel Hill, Chapel Hill, NC 27599, USA
4 Renaissance Computing Institute, University of North Carolina at Chapel Hill, Chapel Hill, NC 27517, USA
5 Siemens AG, Munich, 80333, Germany
6 Kapernikov, Leuven, 3010, Belgium
7 Faculty of Health Sciences, Simon Fraser University, Burnaby, BC V5A 1S6, Canada
8 Community Dentistry and Behavioral Science, University of Florida College of Dentistry, Gainesville, FL 32610, USA
9 Data and Technology, Allen Institute, Seattle, WA 98109, USA
10 Berlin Institute of Health at Charité - Universitätsmedizin Berlin, Berlin 10999, Germany
11 GSK, San Francisco, CA 94080, USA
12 William Harvey Research Institute, Queen Mary University of London, London EC1M 6BQ, United Kingdom
13 Reaction Engineering & Catalyst Development, Leibniz Institute for Catalysis (LIKAT), Rostock, 18059, Germany
14 Scientific Data Division, Lawrence Berkeley National Laboratory, Berkeley, CA 94720, USA
15 Semanticly, Athens, 10563, Greece

16 Knocean Inc., Toronto, Ontario M6P 2T3, Canada
17 Department of Neurology, UCLA, Los Angeles, CA 90095, USA
18 Personalized Health Informatics Group, SIB Swiss Institute of Bioinformatics, Basel 4051, Switzerland
19 Genome Biology Unit, European Molecular Biology Laboratory, Heidelberg, Baden-Württemberg 69117, Germany
20 Schools of Medicine, Public Health, and Nursing, Johns Hopkins University, Baltimore, MD 21287, USA

*Correspondence: smoxon@lbl.gov

# Abstract

Scientific research relies on well-structured, standardized data; however, much of it is stored in formats such as free-text lab notebooks, non-standardized spreadsheets, or data repositories. This lack of structure challenges interoperability, making data integration, validation, and reuse difficult. LinkML (Linked Data Modeling Language) is an open framework that simplifies the process of authoring, validating, and sharing data. LinkML can describe a range of data structures, from flat, list-based models to complex, interrelated, and normalized models that utilize polymorphism and compound inheritance. It offers an approachable syntax that is not tied to any one technical architecture and can be integrated seamlessly with many existing frameworks. The LinkML syntax provides a standard way to describe schemas, classes, and relationships, allowing modelers to build well-defined, stable, and optionally ontology-aligned data structures. Once defined, LinkML schemas may be imported into other LinkML schemas. These key features make LinkML an accessible platform for interdisciplinary collaboration and a reliable way to define and share data semantics.

LinkML helps reduce heterogeneity, complexity, and the proliferation of single-use data models while simultaneously enabling compliance with FAIR (Findable, Accessible, Interoperable, and Reusable) data standards. LinkML has seen increasing adoption in various fields, including biology, chemistry, biomedicine, microbiome research, finance, electrical engineering, transportation, and commercial software development. In short, LinkML makes implicit models explicitly computable and allows data to be standardized at its origin. LinkML documentation and code are available at [linkml.io](https://linkml.io).

## Keywords

# Introduction

Data integration in the sciences is challenging due to heterogeneity, complexity, the proliferation of ad-hoc formats, and poor compliance with FAIR (Findable, Accessible, Interoperable, and Reusable) guidelines [1]. Data is often published in unstructured, text-based formats that lack standardized, reusable structure and element definitions. Without a formal schema to convey the data generator's intent, downstream reuse is difficult, time-consuming, and prone to misinterpretation. Reuse is further hindered by poorly structured formats, missing units or values, unlinked records, and the lack of standardized identifiers or curated relationships. While tools like JSON Schema, relational database schemas, and spreadsheets offer partial solutions, such as structural validation and type enforcement, they lack native support for linking to shared vocabularies or aligning with external data models. These frameworks define data structure but not domain meaning, limiting their ability to ensure that terms, concepts, and relationships used in one dataset are understood in the same way as those in another.

LinkML addresses this gap by providing a unified, schema-driven framework that supports both structural and semantic modeling (**Figure 1**).

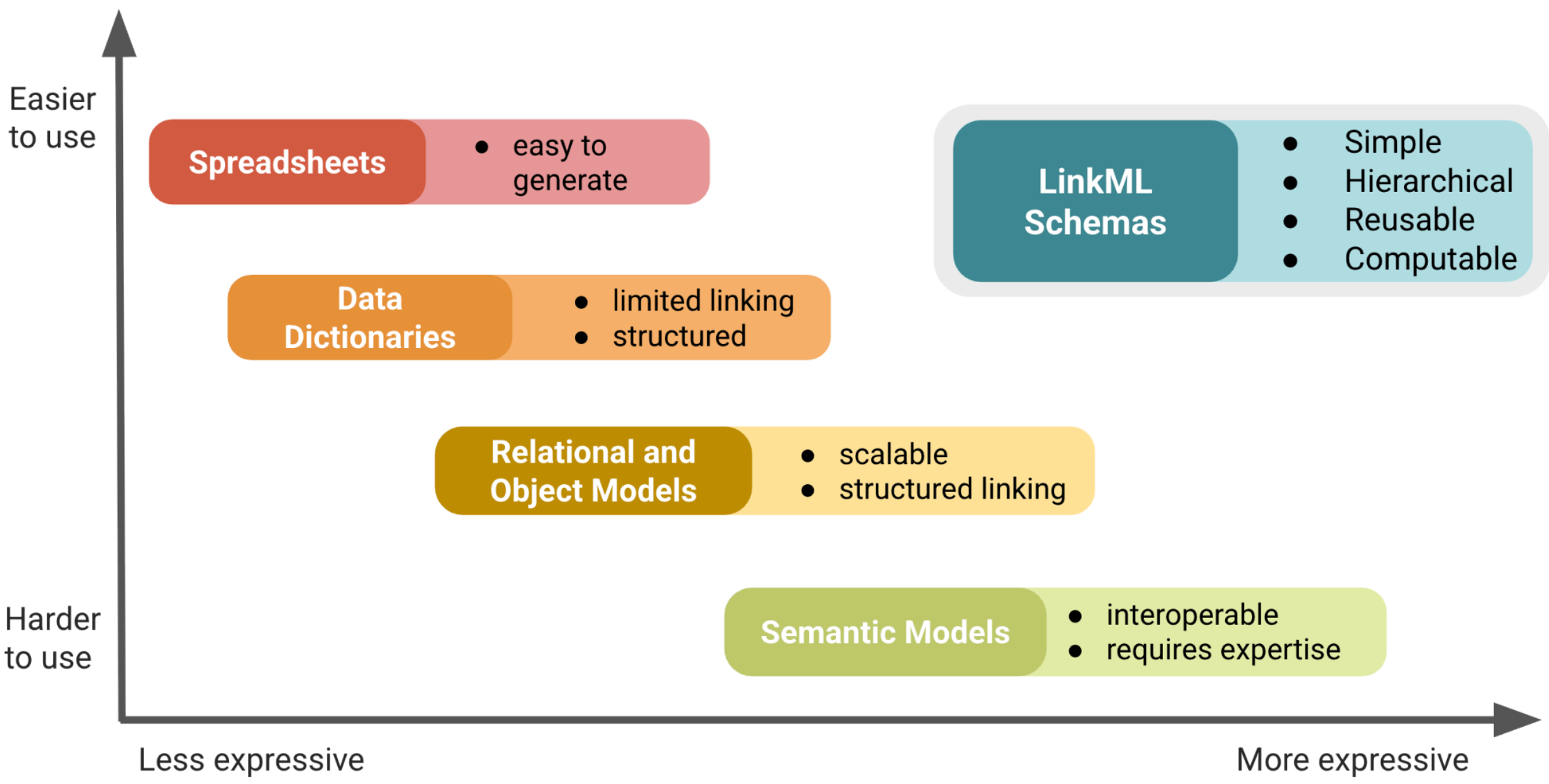

**Figure 1: LinkML is both expressive and easy to use.** LinkML balances ease of authoring with support for semantic clarity, interoperability, and reuse, enabling users to create structured models without requiring extensive technical expertise while allowing for expressive and standards-aligned representations.

## The LinkML Framework

LinkML is an open and extensible data modeling framework that provides a simple and structured way to describe and validate data. As illustrated in **Figure 2**, LinkML facilitates cross-domain collaboration between experts to create shared data models. It includes a schema language (the LinkML metamodel) for defining data elements and their relationships, and tools to convert models to widely used data modelling formalisms like JSON Schema [2], SQL [3], RDF [4], OWL [5], SHACL [6] and Python [7]. Rather than replacing current tools, LinkML augments them with shared semantics, machine-readable documentation, and cross-platform interoperability, making FAIR data easier to produce, validate, and reuse across domains.

LinkML supports a wide range of data structures, from simple spreadsheets and data tables to complex, interlinked models. It helps reduce data inconsistency and complexity while aligning with FAIR data standards [1] by promoting well-defined, persistently identified, and ontology-annotated data structures. It balances ease of use with semantic clarity, offering a powerful alternative to both low-barrier formats like spreadsheets and high-complexity models like RDF/OWL. By combining structured linking, ontology reuse, and human-readable syntax, LinkML enables interdisciplinary teams to build computable, reusable models without sacrificing accessibility or interoperability.

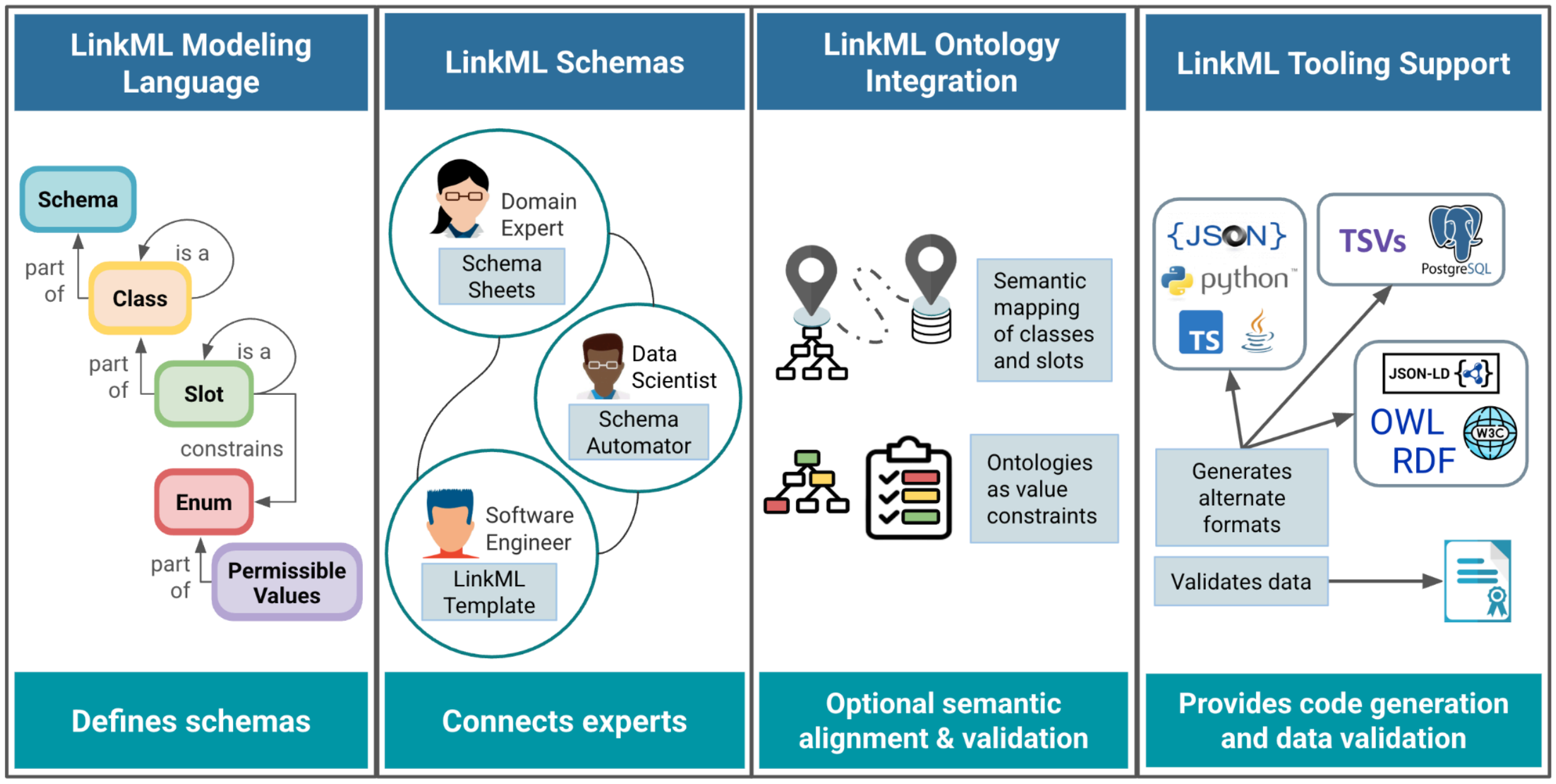

**Figure 2: LinkML framework components.** The LinkML modeling framework provides a flexible way to organize information, author data models, and reuse existing standards like ontologies and linked data frameworks. It can help validate data for schema-conformance, foster expert collaboration, and translate between LinkML and other technical modeling frameworks. The modeling language and tooling can be explored at [8]. **Panel 1**: LinkML schemas consist of hierarchical classes with attributes (slots) that describe them. **Panel 2**: LinkML's data model creation tools are accessible to users with varying levels of expertise. Users can bootstrap a LinkML data model from existing data using Schemasheets or SchemaAutomator, or start with the LinkML Template. **Panel 3:** LinkML can bridge model semantics across formats by reusing knowledge in ontologies, mapping schema elements to ontology terms, and providing translators between representations, such as RDBMS and OWL. **Panel 4**: LinkML generates schemas in several popular formats and validates data.

# Motivating Example

To illustrate LinkML's capabilities, we use biological microbiome sampling as a representative example. Microbiome studies involve collecting samples from diverse environments, such as air, water, organisms, and soil, to perform analyses that characterize these habitats and their associated microbial communities. These analyses may include physicochemical measurements, contaminant testing, and genomic sequencing. Once published, sampling data can be repurposed for many applications, such as building machine-learning models to predict microbial community functions, tracking contamination patterns across ecosystems, constructing knowledge graphs that link samples to biological and chemical data (such as pathways or genomic features), or integrating data into centralized repositories like the National Microbiome Data Collaborative [9].

To appreciate the need for LinkML, consider the typical state of microbiome sampling data. In **Figure 3,** an example data set is shown. The data is in a spreadsheet, but it lacks consistency and clarity. For example, the "dep" field (what is "dep"? It is not defined) has entries that employ a variety of formats and units, so they cannot be easily compared with each other, let alone with other data sources. Terms are not tied to an ontology or terminology, so it's not clear what they cover. For example, can "sand" refer to any type of sand? Do empty cells mean "not specified", "not required", or "mistakenly blank"? And combining latitude and longitude into a single "position" field makes it harder to query or validate locations.

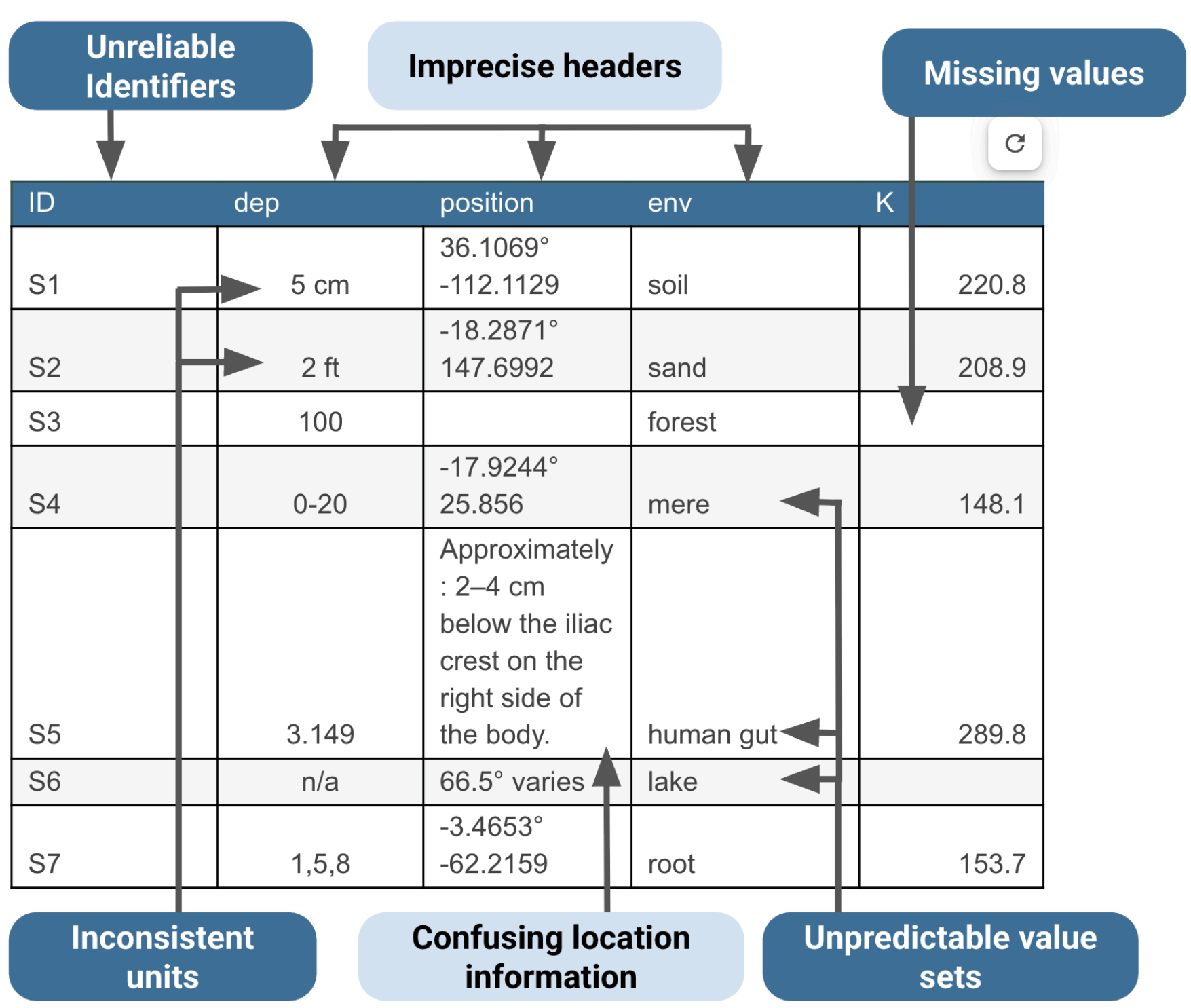

| ID | dep | position | env | K |
|---|---|---|---|---|
| S1 | 5 cm | 36.1069° -112.1129 | soil | 220.8 |
| S2 | 2 ft | -18.2871° 147.6992 | sand | 208.9 |
| S3 | 100 | | forest | |
| S4 | 0-20 | -17.9244° 25.856 | mere | 148.1 |
| S5 | 3.149 | Approximately : 2–4 cm below the iliac crest on the right side of the body. | human gut | 289.8 |
| S6 | n/a | 66.5° varies | lake | |
| S7 | 1,5,8 | -3.4653° -62.2159 | root | 153.7 |

Before LinkML

**Figure 3: An example of sampling data before the use of LinkML.** This "before" example highlights the kinds of inconsistencies and uncertainties researchers face when working with sampling data without a standardized schema. The data is in a spreadsheet, but it is not FAIR. It lacks consistency and clarity and requires human intervention to interpret, harmonize, and parse objectively.

# The LinkML Standard

The LinkML framework provides a flexible data modeling language built around four core element types: schemas, classes, slots, and enumerations. These core language elements allow users to define and organize information in a structured, machine-readable way. A LinkML schema specifies the meaning, relationships, and constraints of a dataset. Within a schema, classes represent conceptual entities (e.g., samples, genes, publications), slots define the attributes or fields associated with these classes, and enumerations (analogous to dropdown menus) constrain slot values to controlled vocabularies. Basic data types, such as strings or integers, are defined by types, and enumerations can be created locally or imported from external ontologies. Building upon these core language constructs, the LinkML metamodel (or standard) is itself expressed as a LinkML schema that precisely defines how schemas, classes, slots, enumerations, and types should be described. This enables LinkML to be self-describing, facilitates tooling, and supports mapping schema elements to existing standards and ontologies **(Table 1)**.

**Table 1**: **LinkML schemas consist of four main elements: classes, slots, types, and enumerations.** In LinkML, classes and slots are defined hierarchically and can be reused across models. Classes can reference each other—for example, by using one class as a range of a slot in another—to represent complex, connected data. Slot values can be constrained by patterns, ranges, or custom data types, and all elements support rich metadata and mappings to external standards, such as ontologies.

| Elements | Description | Examples |
|---|---|---|
| **Schema** | Defines the overall structure of a data model, including metadata, prefixes, and imports. | Sample Schema with prefixes like NMDC |
| **Classes** | Represent entities or concepts in the model, which can be linked to each other via slots. | Sample, Study |
| **Slots** | Attributes or properties of classes, which describe their characteristics and relationships. | latitude, longitude, analysis_type, metagenome_study |
| **Types** | Define data types that slots can hold, such as strings, integers, or enumerations. | string, integer, float, boolean, Class, Enumeration |
| **Enumerations** | Controlled vocabularies to constrain slot values or link to external ontologies. | environment_type: Environment Ontology terms |
| **Mappings** | Link schema elements to external standards or ontologies via URIs. | class_uri: nmdc:Biosample, slot_uri: MIXS:0000009 |

LinkML provides powerful mechanisms for linking knowledge, including (1) connecting model classes within a schema through typed relationships, (2) establishing class hierarchies that support the specialization of concepts, and (3) mapping to existing data standards and ontologies to enhance interoperability. For example, a representative Sample class can be linked to related classes, such as Sample Site or Metagenomic Study, using defined slots that describe relationships like “collected from” or “associated with”. Class hierarchies enable schema developers to define general concepts and then specialize them through subclasses. For instance, an Air Sample Site subclass can inherit all properties of a general Sample Site while adding air-specific details, such as humidity or PM2.5 levels. **Figure 4** further describes the LinkML syntax used in this example. For more detailed information, refer to the LinkML tutorial [10], documentation [11], and metamodel specification [12].

In LinkML, a *mapping* refers to the alignment of schema elements, such as classes, slots, or enumerations, with terms from established ontologies, typically referencing the ontology term's Uniform Resource Identifier (URI). For example, a slot representing the area or type of an environment from which the sample was collected can be constrained to values from the Environment Ontology (ENVO) [13] by defining an enumeration whose permissible values are derived from a subset of ENVO. In exposure and toxicology modeling, this same approach can be used to constrain a slot, such as “exposure stimulus,” to terms from the Exposure Ontology (ECTO) [14]. This enables the reuse of ontology branches, allowing for the validation and interpretation of data based on domain semantics encoded in the ontology. A practical application of this approach can be seen in the NMDC schema [15], which utilizes ENVO terms in enumerations to facilitate the standardization of environmental context annotations in microbiome sample collection data [16].

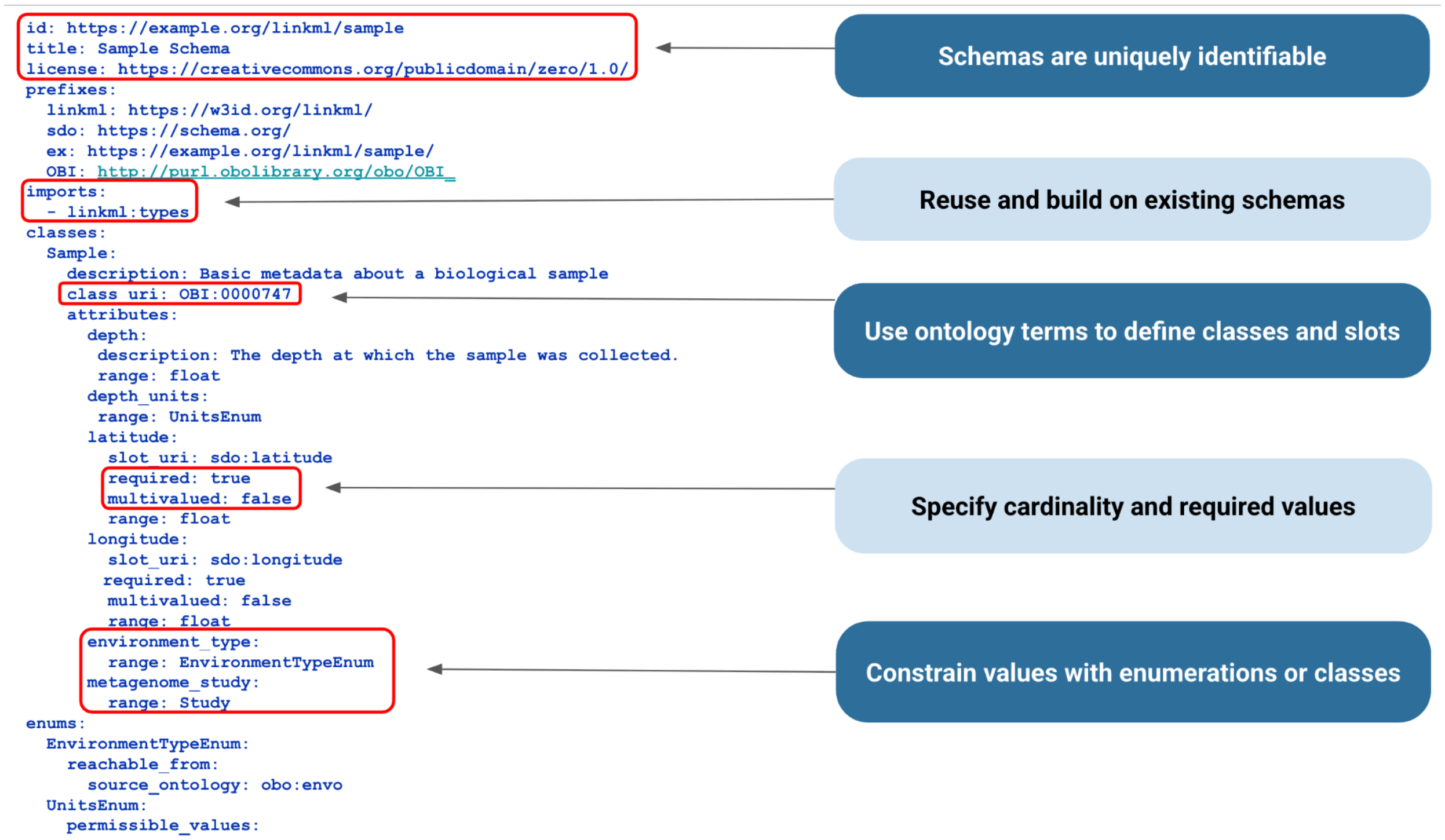

**Figure 4: Example of a LinkML schema.** This snippet shows part of a Sample schema written in YAML [17] using LinkML. The schema begins with a unique identifier in the form of a URL, along with metadata such as `title` and `license`. It defines a set of prefixes to incorporate existing models and uses an “imports” section to bring in other LinkML schemas, promoting reuse instead of reinventing common structures. Existing classes and types are leveraged within the class definition. For instance, the Sample class draws on the Ontology for Biomedical Investigations (OBI) [18] and the `latitude` slot draws on the schema.org [19] definition. Slot properties such as `required` and `multivalued` specify cardinality of the property or slot, while enumerations provide controlled vocabularies that tie allowed values to well-defined semantics.

In LinkML, the attributes of a class are defined using *slots*, which are analogous to database fields or spreadsheet columns. Each slot describes a particular characteristic or property that class instances can have. The slot *range*—such as a string, integer, boolean, or a more complex structure like another LinkML class or an *enumeration*—determines the kind of data that the slot can hold. Slots can also be extensively annotated with metadata, such as textual definitions, usage notes, and mappings to external standards or ontologies. When slots are assigned globally unique identifiers (URIs), they can be referenced reliably in other schemas or external systems. MIxS schema [16], which uses LinkML to standardize the annotation of genomic samples and sequences, organizes related slots into checklists which are structured collections tailored for specific sample types or environmental contexts. Each checklist defines a group of attributes (slots) that describe relevant metadata, such as environmental conditions, host-associated information, and geolocation data. These slots capture the presence of certain data and enforce structure and semantics through their associated types and annotations.

LinkML is designed with reuse and extensibility in mind. Classes, slots, and entire schemas can be imported or extended across schemas, enabling modular modeling and consistent reuse of domain concepts. In addition to these built-in mechanisms, schema alignment, transformation, and evolution can be supported using LinkML-Map [20]. LinkML-Map provides a declarative syntax for expressing changes across schema versions or between related models, including subsetting, renaming, restructuring, and aligning schemas to new standards.

Many real-world projects have adopted LinkML to leverage its core features. For example, the Biolink Model [21] is built using LinkML to define a comprehensive, ontology-aligned schema that connects concepts across diverse biological domains, enabling interoperability among tools, datasets, and knowledge graphs. The NCATS Biomedical Data Translator project [22] builds on this by using the Biolink Model to harmonize data from over 300 sources. The Alliance of Genome Resources (AGR) [23] uses a LinkML schema to integrate data from multiple model organism databases, applying class hierarchies, semantically defined slots, and ontology mappings to handle complex biological relationships and species-specific constraints as their data models evolve.

Returning to the example spreadsheet shown in **Figure 3,** we can see how applying the LinkML schema from **Figure 4** to structure the sample data results in a clearer, more interoperable dataset in **Figure 5.** Using consistent units as dictated by the schema for the depth column eliminates ambiguity, ensures rows are directly comparable, and avoids errors that can arise from interpreting mixed units. LinkML makes it easy to specify identifiers with namespaces and resolvable URIs; these unique and persistent identifiers allow data consumers to unambiguously reference the same entity across time, systems, and studies. The LinkML-structured dataset references environment types using ENVO ontology term identifiers, rather than free text. This standardizes meanings, reduces misinterpretation, and enables computational tools to link samples to broader ecological knowledge.

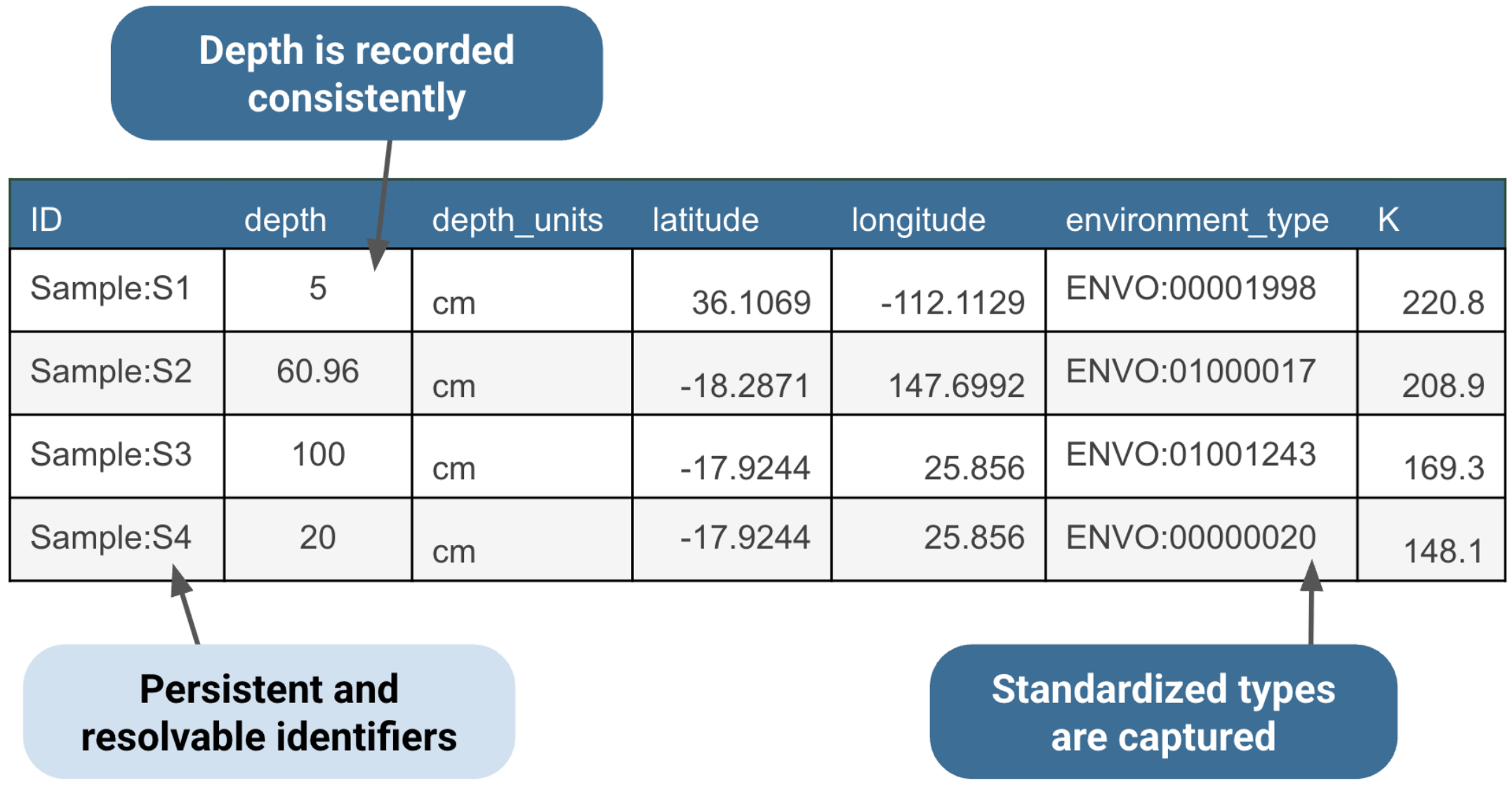

| ID | depth | depth_units | latitude | longitude | environment_type | K |
|---|---|---|---|---|---|---|
| Sample:S1 | 5 | cm | 36.1069 | -112.1129 | ENVO:00001998 | 220.8 |
| Sample:S2 | 60.96 | cm | -18.2871 | 147.6992 | ENVO:01000017 | 208.9 |
| Sample:S3 | 100 | cm | -17.9244 | 25.856 | ENVO:01001243 | 169.3 |
| Sample:S4 | 20 | cm | -17.9244 | 25.856 | ENVO:00000020 | 148.1 |

After LinkML

**Figure 5: After using LinkML to structure the "before" data shown in Figure 3.** A unique and persistent identifier enables each sample to be unambiguously referred to; the depth field is formatted consistently and uses the same units, enabling comparison; the environment type is identified by identifiers from the ENVO environment ontology.

LinkML builds resolvable, persistent identifiers directly into its core modeling framework. The syntax requires explicit identifiers for every schema element and provides built-in mechanisms for prefix management and CURIE-to-URI expansion. LinkML's tooling further supports this by resolving namespace prefixes (via Bioregistry.org and other resolver services), validating CURIEs, and ensuring consistent URI generation. The ecosystem also applies these practices to its own metamodel. Each element of the LinkML metamodel is assigned a persistent identifier that resolves through the w3id permanent identifier service (https://w3id.org/linkml). LinkML's end-to-end use of resolvable identifiers serves as a model for how LinkML schemas can achieve long-term, stable, FAIR-aligned identification. By using LinkML to define and publish both data and schema metadata, researchers can establish clear, shared definitions of entities and their attributes, such as samples, collection sites, or measurement types, ensuring that data is interpretable, comparable, and reusable. This approach benefits not only microbiome studies but also any field that relies on integrating diverse data sources into larger knowledge frameworks.

## LinkML Data Lifecycle

The LinkML data lifecycle is summarized in **Figure 6**, which provides an overview of how LinkML facilitates data modeling from schema creation to validation and integration. Tools in the LinkML framework that support each phase of the data lifecycle are summarized in **Table 2**.

**Table 2**: **Key software tools in the LinkML ecosystem, categorized by the stages of the schema development lifecycle**: Create Schema, Apply Best Practices, Reuse Ontologies, Validate Data, and Manage Schema Evolution. Each tool is briefly described, along with its primary function and the lifecycle stage it supports, illustrating how LinkML enables iterative, structured, and ontology-driven data modeling and validation.

| Tool Name | Description | Lifecycle Stage |
|---|---|---|
| **linkml-project-copier** | A copier-based starter project to scaffold new LinkML schemas with recommended structure. | Create Schema |
| **schemasheets** | Define schemas in spreadsheet format and convert them into LinkML YAML. | Create Schema |
| **schema-automator** | Auto-generates LinkML schemas from source artifacts such as spreadsheets or JSON data. | Create Schema |
| **gen-python, gen-pydantic** | Serializes Python classes that reflect the schema structure, for runtime validation, serialization, and type enforcement | Create Schema |
| **gen-rdf, gen-owl** | Serializes LinkML model as RDF triples using RDFS and SKOS to represent schema structure and semantics | Create Schema |
| **gen-json-schema, gen-jsonld** | Converts the schema into JSON Schema for validating JSON data files, and outputs JSON-LD context files to support semantic annotation and Linked Data compatibility. | Create Schema |
| **linkml-lint** | Validates schema conformance to community best practices using customizable linting rules. | Apply Best Practices |
| **ontogpt** | Suggests ontology mappings and value sets by prompting large language models, such as GPT. | Reuse Ontologies |
| **linkml-validate** | Validates input data against a LinkML schema; supports extensions via custom plugins. | Validate Data |
| **linkml-map** | Supports schema alignment, transformation, and evolution across versions or related models. | Manage Schema Evolution |
| **linkml-store** | Adds a layer of abstraction between data models and underlying technologies, making it easier to migrate or adapt backends without changing the model itself. | Manage Schema Evolution |

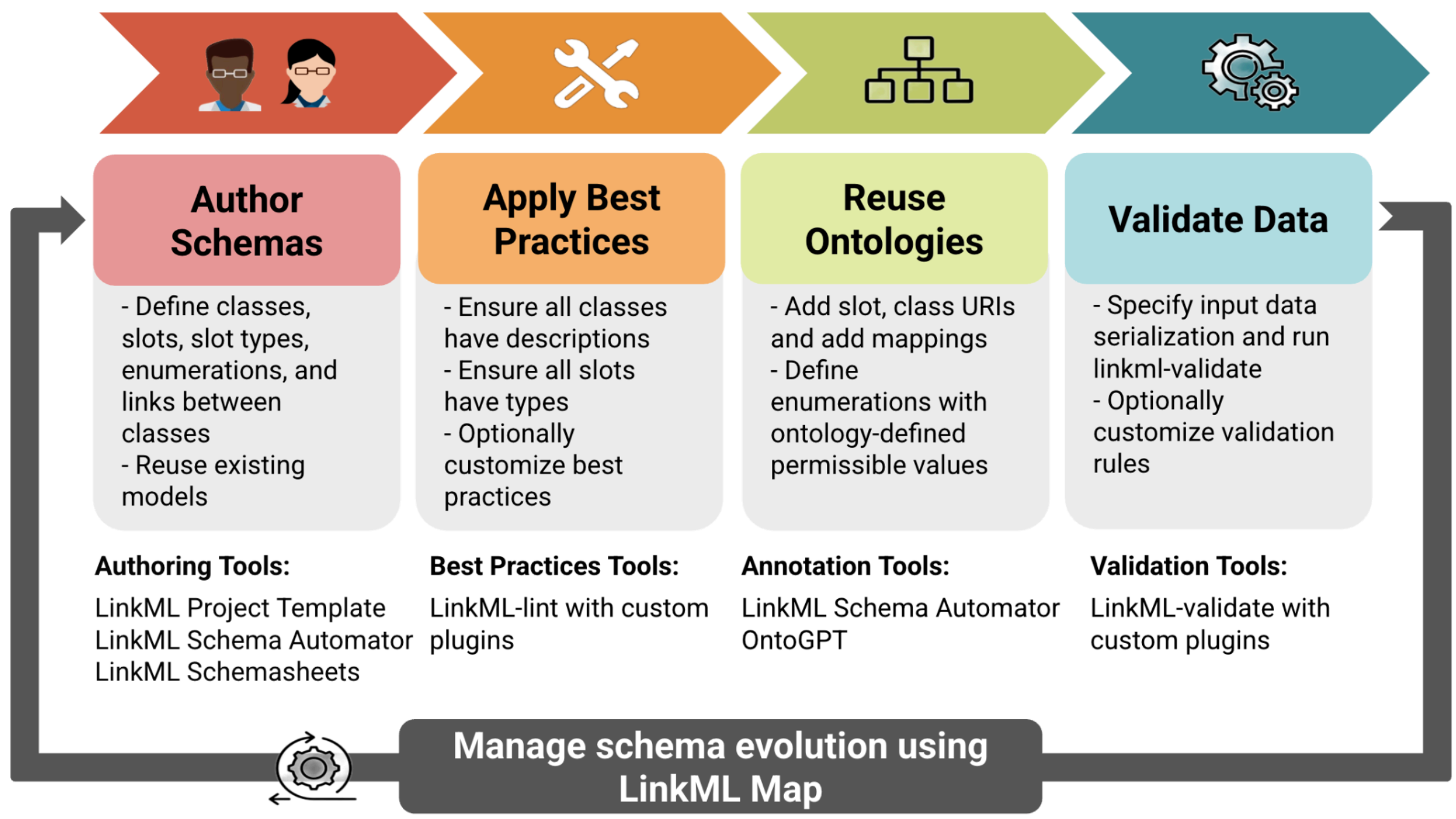

**Figure 6. The five stages of the LinkML data lifecycle.** The LinkML data lifecycle encompasses five main stages: schema creation and reuse, applying best practices, reusing ontologies, validating data, and managing schema evolution. Each stage is supported by a suite of LinkML tools designed to simplify and enhance the process of developing interoperable data models.

## Author Schemas

LinkML aims to be accessible to non-technical users (such as domain experts) as well as data modelers and software developers by providing multiple authoring approaches. Users can directly edit schemas using the default YAML syntax (**Figure 4**) or take advantage of the LinkML project template system, which provides scaffolding for a new LinkML schema project with pre-configured tooling for schema development, documentation and maintenance.

For users who prefer to author in a tabular format, LinkML Schemasheets [24] allow modelers to define classes, slots, and other LinkML elements and schema syntax row by row in a spreadsheet (**Table 3**), which LinkML tools can automatically convert to a YAML representation. Additionally, the LinkML SchemaAutomator tool supports users migrating from other modeling formalisms by converting various sources, such as TSV files, HTML tables, SQL databases, JSON Schemas, and OWL/RDF files, into the LinkML format.

**Table 3. Example schema written in LinkML Schemasheets format using a tabular structure.** LinkML provides tooling to convert spreadsheet-based model representations into LinkML YAML and other model serializations, including JSON Schema, OWL, Python data classes, and Pydantic models. Each row defines either a class (record) or a slot (field) within a class. The `record` or `class` column names the class being defined. The `field` or `slot` column lists attributes (slots) associated with the class. The `multiplicity` or `cardinality` column specifies the allowed number of values, using syntax such as `0..1`(optional), 1 (required), or `0..*` (zero or more). The `required` column indicates whether the field is mandatory. The `range` column defines the expected data type or class of the field.

The `parents` column specifies the superclass or inherited slot. The "`desc`" or `description` column provides a human-readable explanation of the class or field's purpose.

| record | field | multiplicity | required | range | parents | desc |
|---|---|---|---|---|---|---|
| > class | slot | cardinality | required | range | is_a | description |
| > | | | | | | |
| Sample | | | | | | |
| Sample | latitude | 1 | TRUE | decimal | | The latitude of the sample location |
| Sample | longitude | 1 | TRUE | decimal | | The longitude of the sample location |
| Sample | environment_type | 0..1 | | EnvironmentTypeEnum | | The environment type the sample was taken from |
| Study | id | 1 | TRUE | uriorcurie | | The unique identifier of a study |

### Deriving Schema Serializations

LinkML provides extensive schema generation and serialization tools that convert schemas from the default YAML syntax into various formats, including JSON Schema, Python Pydantic [20] classes, SQL DDL, RDF, OWL, ShEx [21], and SHACL [7] (see **Figure 2**). This ability to produce polyglot outputs ensures compatibility across different technical environments. However, conversions are constrained by the expressiveness of the target framework and may result in some loss of information.

## Applying Best Practices

Once developers have created a schema, LinkML encourages them to apply best practices through its linting tool. While LinkML is generally permissive about model style, the linter promotes consistency and clarity by ensuring schemas include clear descriptions, follow naming conventions (e.g., CamelCase for classes, snake_case for slots), specify appropriate slot ranges, and define enumerations with permissible values. Users can customize these checks by configuring their linting rules and tailoring the validation process to their specific requirements.

## Optional Ontology Integration

Using ontologies in LinkML is optional, but can significantly improve semantic clarity and interoperability when needed. For users who do not require formal semantics, LinkML works effectively without the need for ontology integration. However, when precision or alignment with external standards is desired, LinkML provides lightweight and flexible mechanisms to reference ontologies, such as annotating slots or classes with URIs, or declaring mappings (e.g., `skos:exactMatch`, `broadMatch`, `narrowMatch`) to established vocabularies.

A common use case is defining permissible values for a slot using terms from an ontology. For instance, a schema describing microbiome samples might constrain the `environmental_material` slot to values from a specific branch of the Environment Ontology (ENVO), such as terms under "soil" or "marine water." This enables data from different projects to use a shared set of values, while still allowing schema authors to subset or annotate the ontology to meet their local needs. When applied in practice, as in the NMDC submission schema, this approach has helped harmonize environmental context annotations across thousands of samples contributed by multiple labs, enabling federated search, filtering, and comparative analysis.

## Validating Data

Effective data validation is essential to ensuring interoperability. The LinkML validator [25] verifies that data conforms to a specified schema, including those schemas that integrate external ontologies as enumerated value sets or range constraints. Through its plugin-based architecture, the validator supports a variety of validation strategies, including JSON Schema validation, SHACL for semantic validation, and SQL database constraints for ensuring relational data integrity. The LinkML validator allows model designers to declare all constraints within the schema itself, independent of the underlying data storage format, thereby enhancing the flexibility and extensibility of the validation process. This approach also ensures compatibility with existing validation frameworks and provides a mechanism for confirming the correctness of data and compliance with models [26].

## Managing Schema Evolution

The final lifecycle stage addresses the ongoing need to adapt schemas as data standards, best practices, and use cases evolve. For example, a project may decide to merge two fields into one to align with community standards, or restructure parts of a schema to support a new type of analysis. In the Monarch Initiative [27], this challenge is addressed by reusing a focused subset of the broader Biolink Model schema to support research on rare diseases, demonstrating how existing models are often adapted to meet project-specific needs.

LinkML-Map [20] supports schema evolution by allowing modelers to define transformations through a declarative mapping syntax. These transformations can include subsetting a schema, flattening a normalized model for downstream systems, or modifying slot and class definitions, such as changing data types or cardinalities (**Figure 7**). While SSSOM [28] focuses on mapping individual identifiers and ontology terms at the entity level, LinkML-Map extends these principles to structural schema elements, enabling automated, transparent, and maintainable schema evolution. This approach facilitates data migrations across a range of storage systems, including JSON and CSV files, SQL databases, document stores, and graph databases.

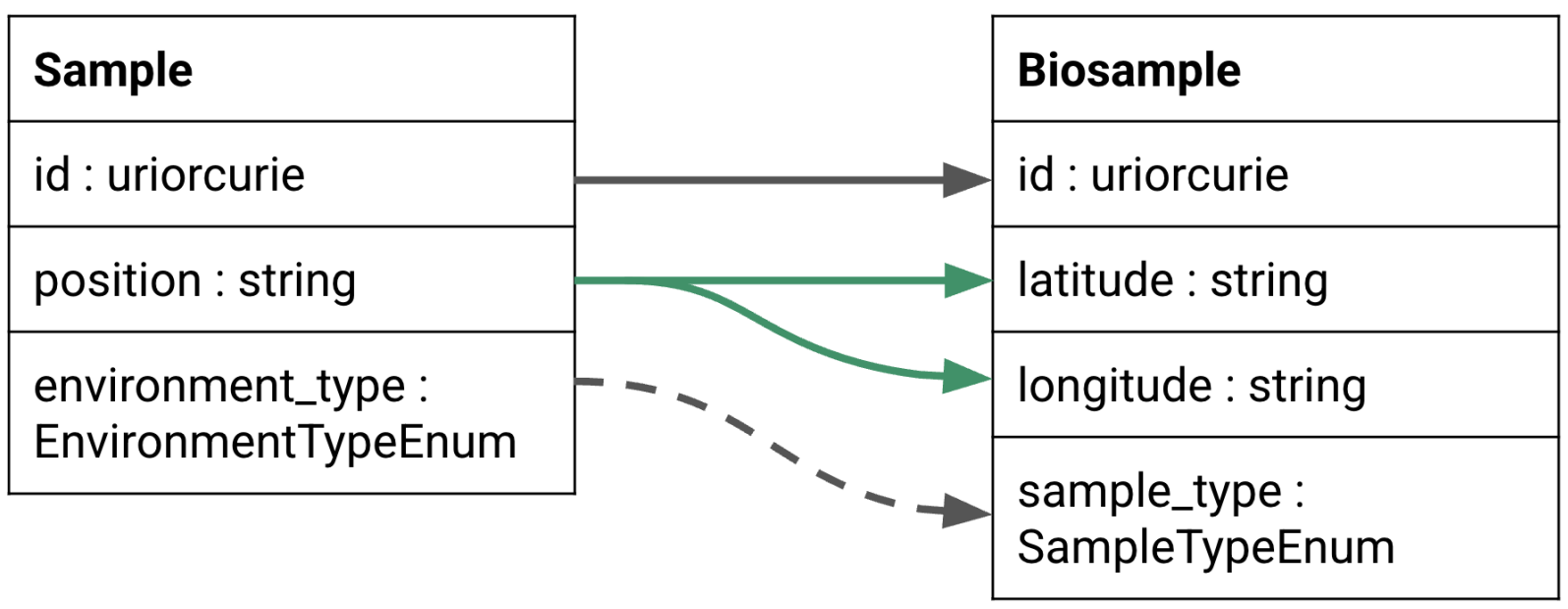

**Figure 7. Example of the LinkML-Map tool tracking changes to a model.** The LinkML-Map software can help manage type and name transformations as well as more extensive merge operations. In the example here, the position field in the source schema is split using LinkML-Map rules into separate fields for latitude and longitude, respectively. Additionally, the environment_type is relabeled to sample_type to align with the terminology used in the target project.

# Documenting and Sharing LinkML Schemas

Effective sharing and deployment of data models are essential for promoting data reuse. However, without standardized mechanisms for distributing models and their metadata, it can be difficult to accurately understand, implement, and reuse data. Model documentation is frequently fragmented, described in disparate and often non-computable formats such as PDFs, plain text documentation pages, SQL dumps, or API annotations, each with varying levels of detail and accessibility. LinkML addresses this challenge by providing a unified, automated documentation system that ensures models are consistently presented in a clear, accessible format (**Figure 8)**. Documentation is deployed to a predictable, web-based location, making it easily discoverable. The schema documentation remains uniform across projects—it is fully indexed, searchable, and visually represented in diagrams. LinkML enables customization of documentation appearance and layout to suit the specific needs of a project, ensuring flexibility without sacrificing consistency.

**Figure 8. LinkML's automated documentation system. Panel A.** Each schema class (e.g., *Sample*) is presented with detailed metadata, including slot descriptions, cardinalities, ranges, and inheritance. **Panel B.** LinkML auto-generates interactive visual schema diagrams to show the relationships between classes, slots, and enumerations (e.g., *environment_type* linked to *EnvironmentTypeEnum*). **Panel C.** The documentation site is automatically deployed to GitHub Pages, ensuring schema updates are reflected without manual publishing steps. **Panel D.** A full-text search interface enables quick discovery of schema components (slots, classes, enums) using labels, descriptions, or identifiers. The documentation also includes URIs, mappings to external standards, and links to the schema source.

# Generative Artificial Intelligence Integrations

LinkML's structured framework complements the growing capabilities of generative artificial intelligence (AI), particularly large language models (LLMs). By providing clearly defined schemas, LinkML serves as a bridge between human engineers, LLMs, and computational tools. The human-readable YAML format that LinkML uses makes schemas easy to author and maintain while providing a structured foundation that LLMs can interact with to enhance data extraction and generation processes. Meanwhile, LinkML makes data AI- and analysis-ready [29] by providing clear, well-defined features and labels, and allowing for the linking and contextualization of scientific data.

For instance, the OntoGPT platform [30] utilizes LinkML schemas to define the data entities and relationships that users want to extract from unstructured text. These schemas are translated into LLM prompts through OntoGPT's LinkML-based code infrastructure, effectively guiding the LLM to generate structured outputs that adhere to the defined schema. This process demonstrates how LinkML enables users to leverage LLMs for tasks that require structured data extraction without sacrificing schema consistency or interoperability. By setting clear expectations about the structure and content of generated outputs, LinkML schemas enhance the usability and reliability of LLM-driven data generation. This alignment between structured schemas and AI-generated content enables a practical, scalable approach to integrating LLMs into existing data ecosystems.

# The LinkML Community

LinkML has a growing community of users and developers building on its open framework. The software is cloned hundreds of times a week, and the main LinkML GitHub repository has over 400 stars as of October 2025. Guided by the O3 (Open Data, Open Code, and Open Infrastructure) principles for sustainable open-source development [31], the LinkML community fosters collaboration through regular community meetings, open discussions, and transparent development practices. Monthly meetings hosted by community members provide a forum for sharing updates, gathering feedback, and exploring diverse use cases. Development primarily takes place in open GitHub repositories, where pull requests, reviews, and discussions drive the ecosystem forward. LinkML is widely adopted, with hundreds of public GitHub projects using it, reflecting its flexibility and broad applicability. Current applications span genomic standards [16], microbiome science [9], clinical genomics [22], model organism research [23], biological pathway modeling [32], knowledge graphs [27], ontology mappings [28], and even large-scale data initiatives like Gaia-X, an effort to build a transparent, federated data infrastructure that also leverages LinkML-based schemas [33].

Community contributions (hundreds of pull requests from over 90 people so far) continue to expand LinkML's capabilities. For example, DataHarmonizer [34], a spreadsheet-like data authoring service, converts LinkML modeling elements into checkable constraints on a web-based data entry form, providing an accessible interface for data entry. Projects such as the NMDC actively use Data Harmonizer. Another example is the community-supported development for representing multidimensional arrays in LinkML [35], which enabled the use of LinkML in projects such as Neurodata Without Borders.

The LinkML Registry [36] serves as a curated collection of schemas created using LinkML, similar in purpose to the OBO Foundry's [37] approach to organizing interoperable ontologies. The registry provides diverse examples that can be used and adapted for new data modeling projects.

# Conclusions

While expressive frameworks such as the Resource Description Framework (RDF) and Labeled Property Graphs (LPGs) offer semantically rich ways to represent data, most datasets are still shared in simpler formats like SQL dumps, CSV/TSV files, and spreadsheets.These define structure but not domain meaning, making it challenging to ensure that terms and relationships are interpreted consistently across datasets. They persist because they are easy to use, but they fall short when it comes to interoperability, semantic clarity, and model reuse. Community-driven frameworks such as ISA-Tools for experimental metadata [38] and Frictionless Data [39] provide valuable solutions for tabular data and predefined schemas. Formats like Croissant [40] standardize dataset-level metadata. Similarly, Maggot [41] offers an extensible schema, web interface, and storage-linked metadata management for datasets, aligned with data management plans. LinkML complements these approaches by enabling self-authored, customizable tabular schemas, hierarchical classes, ontology integration, multi-format data exchange, and automated code generation. But LinkML goes further: it models not only the metadata about a dataset but also the structure and relationships among the entities within it, enabling consistent representation of both datasets and their contents [42].

LinkML fills an important gap by offering a modeling language that is intuitive for domain experts yet expressive enough for ontology-aware, structured data integration. It provides a single source of truth: models are defined once and translated consistently into various formats, including JSON, YAML, RDF, Python, and SQL. Its metamodel supports human-readable and machine-interpretable definitions of classes, properties, relationships, and constraints, making it well-suited for heterogeneous systems. As an open-source framework with an active community, LinkML continues to evolve alongside emerging standards. By enabling schema-driven development across diverse tooling ecosystems, LinkML helps teams collaboratively build models that are reusable, computable, and aligned with modern data-sharing goals. Rather than forcing a choice between simplicity and semantic rigor, LinkML offers both, making it a pragmatic solution for projects and communities aiming to make their data more reusable and interoperable.

# Availability of supporting source code and requirements

Project name: LinkML
Project home page: https://linkml.io (documentation), https://github.com/linkml (source code)
Operating system(s): Platform independent
Programming language: Python and others
Other requirements: Python >= 3.9 as of 2025-06-27
License: The LinkML model and specification is provided under a CC-0 license waiver.
Supporting tooling is available as a mixture of BSD-3 and Apache 2.0.
Release archives (on Zenodo): https://doi.org/10.5281/zenodo.5703670

RRID: RRID:SCR_027188

# Abbreviations

RDF: Resource Description Framework
RDBMS: Relational Database Management System
LPG: Labeled Property Graph
O3: Open Data, Open Code, and Open Infrastructure
NMDC: National Microbiome Data Collaborative
URI: Uniform Resource Identifier
ENVO: Environment Ontology
ECTO: Exposure Ontology
MIxS: Minimal Information about any Sequence
NCATS: National Center for Advancing Translational Science
SSOM: Simple Standard for Sharing Ontology Mappings
LLM: Large Language Model
AI: Artificial Intelligence
OBO: Open Biological and Biomedical Ontologies

# Competing Interests

The authors declare that they have no competing interests.

# Funding

US Department of Energy, Office of Basic Energy Sciences, DE-AC02-05CH11231, (SATM, SP, NLH, JHC, AJI, PK, MAM, CJM);
National Human Genome Research Institute, Center of Excellence in Genome Sciences, RM1 HG010860, (NLH, JHC, MAH, NM, JOBJ, KS, CJM, SG);
Chan Zuckerberg Initiative / Wellcome Trust, Essential Open Source Software for Science program (EOSS Cycle 6), 313291/Z/24/Z, (MHB, CC, CK, KS, SG);
National Institutes of Health, National Institute of Mental Health, 4DP2MH129986, JLS;
National Institutes of Health, National Institute of Neurological Disorders and Stroke, U24NS120057, RL;
National Institutes of Health, National Institute of Mental Health, 1U24MH130919-01, TF;
National Institutes of Health, National Institute of Dental and Craniofacial Research, 1U01DE033978-01, WDD;
US Department of Agriculture, Non-Assistance Cooperative Agreement, 58-8040-8-014-F, DMD;
Genome Canada, Bioinformatics and Computational Biology, 286GET, DMD;
National Institutes of Health, National Center for Advancing Translational Sciences, 1OT2TR005712-01, CB;
National Cancer Institute, Center for Cancer Data Harmonization, 19X077Q HHSN261201500003I, GV;

# Authors' contributions

**Conceptualization:** CJM, DRU, HS, MAH, OR, SATM
**Funding acquisition:** CJM, MAH, NLH, SATM, MHB
**Project administration:** NLH, SATM, SG
**Resources:** CB, CJM, DRU, JHC, KS, NLH, SATM, SP, WAV
**Software:** AG, AJI, CC, CJM, CK, DL, DMD, DRU, FD, GV, HH, HS, JLS, JOBJ, KS, MAM, MK, NM, PK, RL, SATM, SCC, SP, TF, WMAMV, WDD
**Supervision:** CGC, CJM, KS, MAH, MHB, SATM
**Validation:** AG, CB, CC, CGC, DRU, FD, GV, HS, JAO, JHC, MAM, MHB, NM, OR, RL, SATM, SCC, TF, WMAMV
**Writing – original draft:** CJM, NLH, SAM
**Writing – review & editing:** CJM, DL, DRU, JFY, JHC, MAH, MHB, NLH, NM, SATM, SG

**LinkML Community Members (listed in alphabetical order)**: Richard M. Bruskiewich, Seth Carbon, Eric Cavanna, John-Marc Chandonia, Shreyas Cholia, Ben Dichter, Emiley A. Eloe-Fadrosh, Vincent Emonet, Shahim Essaid, James A. Fellows Yates, Joseph Flack, Satrajit S. Ghosh, Damien Goutte-Gattat, Dorota Jarecka, Dazhi Jiao, Marcin P. Joachimiak, Vlad Korolev, Volodymyr Lapkin, Noel McLoughlin, Sierra D. Miller, Michael Milton, Josh Moore, Moni Munoz-Torres, B. Nolan Nichols, Tim Putman, Justin T. Reese, Victoria Savage, Philip Stroemert, Jeremy Teoh, Anne Thessen, Isaac To, Puja Trivedi, Vincent Vialard, Trish Whetzel

# Acknowledgements

We wish to thank the entire LinkML community, including those who are not listed as masthead or consortial authors but who have made contributions to LinkML.